\documentclass[aps,prb,twocolumn,superscriptaddress,floatfix]{revtex4}

\usepackage{bm}
\usepackage{amsmath}
\usepackage{amssymb}
\usepackage{graphicx}
\usepackage{color}
\usepackage{soul,xcolor}
\setstcolor{blue}
\usepackage{ulem}
\usepackage{cancel}

\def\sig{{\mbox{\boldmath{$\sigma$}}}}

\makeatletter
\@ifundefined{textcolor}{}
{%
 \definecolor{BLACK}{gray}{0}
 \definecolor{WHITE}{gray}{1}
 \definecolor{RED}{rgb}{1,0,0}
 \definecolor{GREEN}{rgb}{0,1,0}
 \definecolor{BLUE}{rgb}{0,0,1}
 \definecolor{CYAN}{cmyk}{1,0,0,0}
 \definecolor{MAGENTA}{cmyk}{0,1,0,0}
 \definecolor{YELLOW}{cmyk}{0,0,1,0}
\definecolor{ORANGE}{rgb}{1,0,1} }

\@ifundefined{definecolor}
 {\@ifundefined{definecolor}
 {\@ifundefined{definecolor}
 {\@ifundefined{definecolor}
 {\@ifundefined{definecolor}
 {\usepackage{color}}{}
}{}
}{}
}{}
}{}

\usepackage{epsfig}

\def\nab{{\mbox{\boldmath{$\nabla$}}}}
\def\sig{{\mbox{\boldmath{$\sigma$}}}}

\def\b0{{\bf{0}}}

\newcommand{\up}{\uparrow}
\newcommand{\down}{\downarrow}

\begin{document}

\title{Effects of different lead magnetizations on the Datta-Das spin field-effect transistor}

\author{A. Aharony}
\email{aaharonyaa@gmail.com}
\affiliation{Raymond and Beverly Sackler School of Physics and Astronomy, Tel Aviv University, Tel Aviv 69978, Israel}
\affiliation{Physics Department, Ben Gurion University, Beer Sheva 84105, Israel}
%
\author{O. Entin-Wohlman}
\affiliation{Raymond and Beverly Sackler School of Physics and Astronomy, Tel Aviv University, Tel Aviv 69978, Israel}
%

\author{K. Sarkar}
\email{sarkark@post.bgu.ac.il}
\affiliation{Physics Department, Ben Gurion University, Beer Sheva 84105, Israel}

\author{R. I. Shekhter}
\affiliation{Department of Physics, University of Gothenburg, SE-412
96 G{\" o}teborg, Sweden}

\author{M. Jonson}
\affiliation{Department of Physics, University of Gothenburg, SE-412
96 G{\" o}teborg, Sweden}

\date{\today}

\begin{abstract}

A Datta-Das spin field effect transistor is built of a one-dimensional weak link, with Rashba spin orbit interactions (SOI), which connects two  magnetized reservoirs. The particle and spin currents between the two reservoirs are calculated to lowest order in the tunneling through the weak link and in the wide-band approximation, with emphasis on their dependence on the origins of the `bare' magnetizations in the reservoirs.
The SOI is found to generate magnetization components in each reservoir, which rotate in the plane of the electric field (generating the SOI) and the weak link, only if the `bare' magnetization of the other reservoir has a non-zero component in that plane. 
The SOI affects the charge current only if both reservoirs are polarized. 
The  charge current is conserved, but the transverse rotating magnetization current is not conserved since the SOI in the weak link generates extra spin polarizations which are injected into the reservoirs.

 \end{abstract}

\pacs{72.25.Hg,72.25.Rb}

\maketitle

\section{Introduction}
\label{INTR}

Spintronics takes advantage of the electronic spins in designing a variety of applications, including   giant magnetoresistance sensing, quantum computing,  and quantum-information processing. \cite{wolf,zutic} A promising approach for the latter  exploits mobile qubits, which carry the quantum information via the spin polarization of the moving electrons. The spins of mobile electrons can be manipulated  by  the spin-orbit interaction (SOI), which  causes  the spin of an electron moving through a spin-orbit active material (e.g.,  semiconductor heterostructures \cite{Kohda})  to rotate around an effective magnetic field. \cite{winkler,manchon} In the particular case of the Rashba SOI, \cite{rashba} both  the direction of the rotation axis and the amount of rotation can be tuned by  gate voltages. \cite{Nitta,Sato,Beukman,comDres}

In the simplest device, electrons move between two large electronic reservoirs, via a mesoscopic link.
Research in  this  direction was  enhanced following the proposal by
Datta and Das,  \cite{datta}
of a spin field-effect transistor (SFET) based on magnetic reservoirs. When the conduction electrons in the reservoirs are fully spin polarized in opposite directions and when the link is not spin-orbit active then the current between them is completely blocked. However, the existence of the SOI on the link can rotate the electronic spins, and open the blocking. Alternatively, when the reservoir polarizations are parallel the current between them may be blocked when the SOI rotates the spins by $180^o$. Although the literature contains many papers on possible realizations of the Datta-Das SFET,\cite{DDhistory,recent,zutic1} most of these consider the Datta-das SFET with fully polarized conduction electrons in the reservoirs, and do not discuss the dependence of the particle and spin currents on the details of the (possibly partial) reservoir magnetizations. This analysis is presented below. Reference \onlinecite{flensberg} does discuss the more general case, using a different approach than ours, but they do not give explicit expressions for the spin currents or for the different mechanisms that cause the reservoir magnetizations.

\begin{figure}[htp]
\includegraphics[width=8cm]{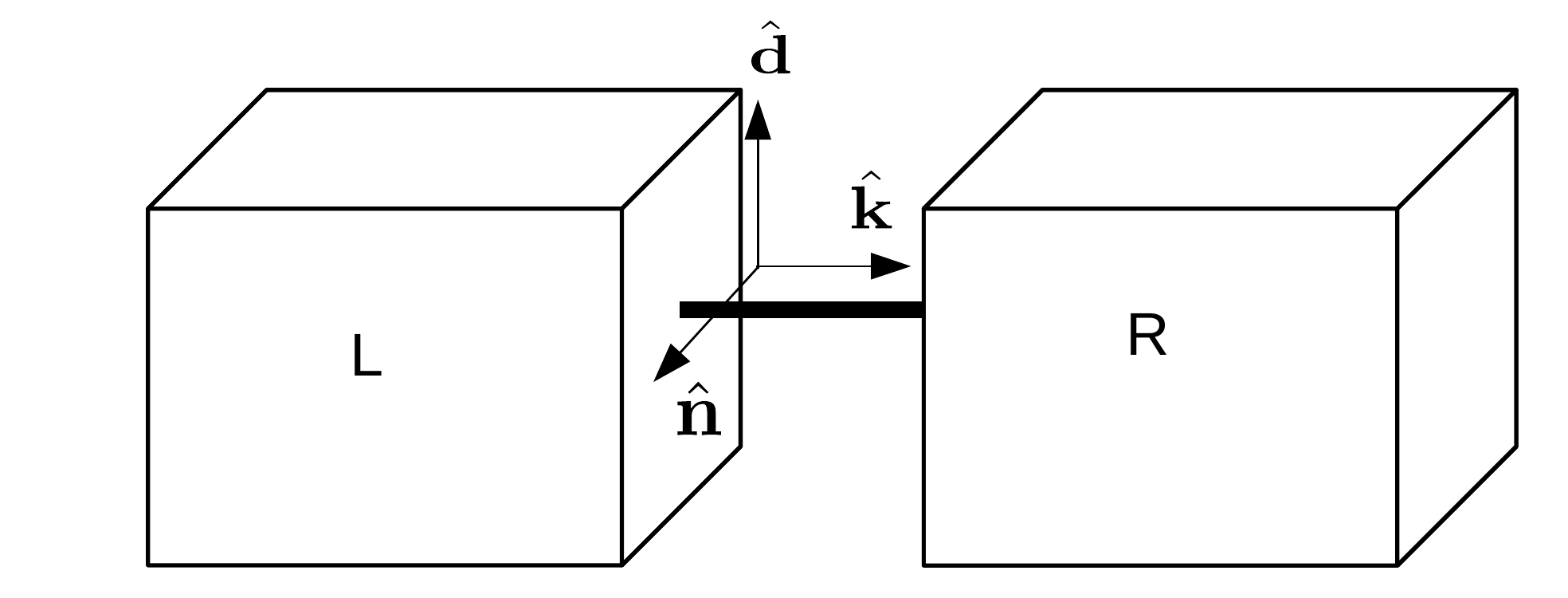}
\caption{
A spin-orbit active weak-link wire (thick line) connecting  two polarized reservoirs, $L$ and $R$. The momentum of the electron is ${\bf k}$, the electric field which generates the spin-orbit interaction is along $\hat{\bf n}$
and the effective magnetic field due to the spin-orbit interaction is along $\hat{\bf d}$.
 }
\label{f1}
\end{figure}

When the link is spin-orbit active, the single-channel transmission  is described by a $2\times 2$ matrix in spin Hilbert space.  Since this matrix is proportional to the unit matrix when time-reversal symmetry is obeyed,  \cite{bardarson}  spin splitting cannot be achieved with SOI alone. One possible way to break time-reversal symmetry is  by applying  a magnetic field. Indeed,  several  proposed devices  utilize an orbital Aharonov-Bohm magnetic flux, which penetrates loops of interferometers to achieve spin splitting, \cite{lyanda,us,Saarikoski} via
the  interference of the spinor wave functions in the two branches of the loop. 
Here we  analyze an even simpler geometry: the two reservoirs are connected by a {\it single} (weak link) spin-orbit active wire, see Fig. \ref{f1}. In a recent paper, \cite{Aharony_2018} the time reversal symmetry was broken by a Zeeman energy gained from an external  magnetic  field acting on the  wire. For certain directions of
this field, both the charge and the spin conductances of the device were found to exhibit oscillations with the length of the wire.
Alternatively, our earlier papers \cite{Shekhter_2013,Shekhter_2014} considered breaking the time-reversal symmetry by using spin-polarized reservoirs, in the context of mechanically-controlled bendable wires.

In this paper we present a systematic analysis of spin splitting through a single straight spin-orbit active wire, with various configurations of polarized reservoirs. After deriving the general expressions for the particle and spin currents through the weak spin-orbit active link, we consider two basic scenarios. In one, all the electrons have spin-independent chemical potentials (and therefore spin-independent Fermi distributions), and the reservoir magnetizations result only from the spin-dependence of the densities of states in the reservoirs. Special cases include paramagnets, where the spin dependence of the densities of states result from external magnetic fields, and ferromagnetic half metals, where each reservoir allows only for one direction of the electronic spins. The latter case exemplifies the Datta-Das original idea, with fully polarized reservoirs (Datta and Das considered ferromagnetic materials like iron, not necessarily half metals). In a second scenario,
a microwave radiation  can generate  non-equilibrium reservoir states which can be described by spin-dependent chemical potentials.

After describing our model in Sec. \ref{model}, Sec. \ref{response} presents the general expressions for the particle and spin currents in the reservoirs. The details of the calculation, which is done to lowest order in the tunneling energies through the weak link and with the wide band approximation, are given in Appendix A. These general results are then applied in Sec. \ref{res} to the special case in which the spin-dependence of the tunneling energies results only from the SOI, so that they are described by a unitary $2\times 2$ matrix. These results are then summarized in Sec. \ref{sum}.

%
%


\section{Model}\label{model}
In this section we describe the model. A spin-orbit active weak-link is
connected to two reservoirs as shown in Fig.~\ref{f1}. The electrons in these
reservoirs are in general spin polarized, as discussed in Sec. \ref{INTR}.
The Hamiltonian of the system is
\begin{align}
{\cal H}={\cal H}^{}_{\rm link}+{\cal H}^{}_{\rm leads}+{\cal H}^{}_{\rm tun}\ .
\end{align}
The Hamiltonian in the weak link, ${\cal H}^{}_{\rm link}$,
contains the kinetic energy and
the linear Rashba spin-orbit interaction (SOI), \cite{Shekhter_2013,Shekhter_2014}
\begin{align}
{\cal H}^{}_{\rm link}=-\frac{1}{2m^{\ast}_{}}\nab^2_{}
-i\frac{\widetilde{k}^{}_{\rm so}}{m^{\ast}_{}} \hat{\bf n}\cdot[\sig\times\nab]\ ,
\label{ham}
\end{align}
where $\sig$ is the vector of Pauli matrices,  $m^{\ast}$ is the electron effective mass, 
$\hat{\bf n}$ is a unit vector along the electric field which
causes the SOI, and we adopt units in which $\hbar=1$. The net strength of the SOI interaction (in momentum
units) 
is denoted $\widetilde{k}_{ \rm so}$. On the one-dimensional link, the electron has a momentum ${\bf k}$ parallel to the link, and therefore
we can write
\begin{align}
{\cal H}^{}_{\rm link}&=\frac{{\bf k}^2_{}}{2m^{\ast}_{}}
-\frac{\widetilde{k}^{}_{\rm so}}{m^{\ast}_{}} [\hat{\bf n}\times{\bf k}]\cdot\sig
=\frac{{\bf k}^2_{}}{2m^{\ast}_{}}-{\bf H}^{}_{\rm so}\cdot\sig\ ,
\label{ham1}
\end{align}
where ${\bf H}^{}_{\rm so}=(k k^{}_{\rm so}/m^{\ast})\hat{\bf d}$, with $\hat{\bf d}=\hat{\bf n}\times\hat{\bf k}/|\hat{\bf n}\times\hat{\bf k}|$ and $k^{}_{\rm so}=\widetilde{k}^{}_{\rm so}|\hat{\bf n}\times\hat{\bf k}|$. Below we assume for simplicity that $\hat{\bf n}$ is perpendicular to the wire, hence
\begin{align}
\hat{\bf d}=\hat{\bf n}\times\hat{\bf k}\ ,
\label{ddd}
\end{align}
and
$\widetilde{k}^{}_{\rm so}=k^{}_{\rm so}$.

The Hamiltonian of the left ($L$) and right ($R$) leads is ${\cal H}^{}_{\rm leads}=
\sum_{\alpha=L,R}{\cal H}^{\alpha}_{\rm lead}$, with
\begin{align}
{\cal H}^{\alpha=L(R)}_{\rm lead}=\sum_{{\bf k} ({\bf p}),\sigma}\epsilon^{}_{ k(p)}c^{\dagger}_{{\bf k}({\bf p})\sigma}c^{}_{{\bf k}({\bf p})\sigma}
\ ,
\label{Hl}
\end{align}
where $c^{\dagger}_{{\bf k}({\bf p})\sigma}$ ($c^{}_{{\bf k}({\bf p})\sigma}$) creates (annihilates) an electron with momentum ${\bf k}({\bf p})$ and spin $\sigma$ in the left (right) lead. The energies $\epsilon^{}_{k(p)}$ denote only the kinetic energies of the electrons. The magnetization of the electrons in the reservoirs is introduced below via the spin-dependence of the densities of state or of the chemical potentials.

In the following we assume that the electrons in both reservoirs are polarized along the same quantization axis, denoted by $\hat{\bf z}$, and $\sigma=+1,-1=\uparrow,\downarrow$ indicate the electronic spin eigenvalues $\sigma\hbar/2$ along that direction. Below we consider the particle current and  magnetization rates in the reservoirs for several choices of the direction $\hat{\bf z}$.  A priori, $\hat{\bf z}$ has nothing in common with the axes shown in Fig. 1.
The electronic spin polarization (or magnetization) in the reservoirs can originate from several sources. First, consider paramagnetic reservoirs. A Zeeman field $B$ along $\hat{\bf z}$  generates spin-dependent energies, $\epsilon^{}_{k(p)\sigma}=\epsilon^{}_{k(p)}+ g\mu^{}_B B\sigma$. Since the electronic densities of states count the states with different momenta (${\bf k}$ or ${\bf p}$), this results in spin-dependent  densities of states at a given total energy $\epsilon$, ${\cal N}^{}_{L(R)\sigma}(\epsilon)={\cal N}^{}_{L(R)0}(\epsilon- g\mu^{}_B B\sigma)$. \cite{AM} 
At low temperatures, ordered ferromagnetic reservoirs may have electrons with only one spin direction, e.g. with only ${\cal N}^{}_{L(R)\uparrow}$. The latter case exemplifies the original Datta-Das model.

Alternatively, consider non-magnetic metallic reservoirs. A non-equilibrium magnetization can be created in the reservoirs by, e.g.,  a microwave radiation which induces a spin-flip assisted absorption of photons. Under the simplified assumption that all the electronic relaxation processes except that of the spins have occurred we may characterize the spin-up and spin-down electrons by Fermi distributions which contain spin-dependent chemical potentials,
\cite{Shekhter_2014}
\begin{align}
f^{}_{L\sigma}(\epsilon)=1/[e^{\beta^{}_{}(\epsilon-\mu^{}_{L\sigma})}+1]\ , \nonumber \\
f^{}_{R\sigma}(\epsilon)=1/[e^{\beta^{}_{}(\epsilon-\mu^{}_{R\sigma})}+1]\ ,
\label{Fermi}
\end{align}
where $\beta^{}=(k^{}_BT)^{-1}$  is the inverse  temperature and $\mu^{}_{L(R)\sigma}=\mu_{L(R)}+\sigma U_{L(R)}$ is the spin dependent chemical potential at the left (right) lead, and where  the spins in both reservoirs are quantized along the same $\hat{\bf z}-$axis.
In contrast to the paramagnetic metals considered above, such a distribution is essentially a non-equilibrium one: it is supported by an external pumping of spin, and when the pumping stops it eventually decays due to  spin-relaxation processes.

Defining
\begin{align}
N^{L(R)}_{0\sigma}=\sum_{k(p)}f^{}_{L\sigma}(\epsilon_{k(p)\sigma})=\int d\epsilon{\cal N}^{}_{L(R)\sigma}(\epsilon)f^{}_{L(R)\sigma}(\epsilon)\ ,
\label{NLRs}
\end{align}
the `bare' total electron number (or density) in the left (right) lead and the magnetization there are
\begin{align}
N^{L(R)}_0&=N^{L(R)}_{0\uparrow}+N^{L(R)}_{0\downarrow}\ ,\nonumber\\
{\bf M}^{L(R)}_{0}&=\big(N^{L(R)}_{0\uparrow}-N^{L(R)}_{0\downarrow}\big)\hat{\bf z}\ .
\label{NM}
\end{align}
These `bare' quantities pertain to the reservoirs alone, i.e., in the absence of the weak link.

 Tunneling between the leads and the weak link is described by
\begin{align}
{\cal H}^{}_{\rm tun}=
\sum_{{\bf k},{\bf p},\sigma,\sigma'}([V^{}_{{\bf k}{\bf p}}]^{}_{\sigma\sigma'}c^{\dagger}_{{\bf k}\sigma}c^{}_{{\bf p}\sigma'}+[V^{\ast}_{{\bf k}{\bf p}}]^{}_{\sigma\sigma'}c^{\dagger}_{{\bf p}\sigma'}c^{}_{{\bf k}\sigma})\ ,
\label{Htun}
\end{align}
where $[V^{}_{{\bf k}{\bf p}}]^{}_{\sigma\sigma'}$ is the tunneling amplitude from the state with momentum ${\bf p}$ and spin $\sigma'$ in the right lead to the state with momentum ${\bf k}$ and spin $\sigma$ in the left one. This amplitude, the key ingredient of our approach,  is proportional to the spin-dependent propagator  connecting  the two states.  \cite{Aharony_2018}

For the Hamiltonian (\ref{ham1}), without an external magnetic field in the weak link, the $2\times2$ matrix $V^{}_{{\bf k}{\bf p}}$ is unitary,
\begin{align}
V^{}_{{\bf k}{\bf p}}=Je^{-i\alpha\hat{\bf d}\cdot\sig}\ ,
\label{VV}
\end{align}
with \cite{Aharony_2018}
\begin{align}
\alpha=k^{}_{\rm so}\ell\ ,
\label{Alp}
\end{align}
where $\ell$ is the length of the wire. Below we concentrate on this special case, and leave more complicated examples for a future publication.

\section{Response to different reservoirs polarization}\label{response}

We now calculate the steady state changes of the electronic particle number and the total magnetization  on the reservoirs, due to the tunneling between them. The rate of change of these quantities (sometimes called `the spin current') in the left lead are determined by
\begin{align}
&R^{L}_{\sigma\sigma'}=\frac{d}{dt}\sum_{\bf k}\langle c^{\dagger}_{{\bf k}\sigma}c^{}_{{\bf k}\sigma'}\rangle=i\sum_{\bf k}\langle[{\cal H}, c^{\dagger}_{{\bf k}\sigma}c^{}_{{\bf k}\sigma'}]\rangle\nonumber\\
&=i\sum_{{\bf k},{\bf p},\sigma^{}_{1}}\langle [V^{\ast}_{{\bf k}{\bf p}}]^{}_{\sigma\sigma^{}_{1}} c^{\dagger}_{{\bf p}\sigma^{}_{1}}c^{}_{{\bf k}\sigma'}-
[V^{}_{{\bf k}{\bf p}}]^{}_{\sigma'\sigma^{}_{1}}c^{\dagger}_{{\bf k}\sigma}c^{}_{{\bf p}\sigma^{}_{1}}
\rangle\ ,
\label{RA}
\end{align}
where the angular brackets indicate a quantum average.

The total rate of change of the particle number (sometimes called `the particle current') into the left lead is then
\begin{align}
I^{L}_{}=\sum_{\sigma}R^{L}_{\sigma\sigma}=R^{L}_{\uparrow\uparrow}+R^{L}_{\downarrow\downarrow}\ .
\label{IL}
\end{align}
 The magnetization in the left lead is
\begin{align}
{\bf M}^{L}_{}=\sum_{{\bf k},\sigma,\sigma^{\prime}}\langle c^{\dagger}_{{\bf k}\sigma}[\sig]^{}_{\sigma\sigma'}c^{}_{{\bf k}\sigma'}\rangle\ .
\end{align}
Therefore, the rate of change of this magnetization can be written as
\begin{align}
&\dot{\bf M}^{L}_{}=\sum_{\sigma,\sigma'}R^{L}_{\sigma\sigma'}[\sig]^{}_{\sigma\sigma'}\nonumber \\ =&R^{L}_{\uparrow\uparrow}[\sig]^{}_{\uparrow\uparrow}+R^{L}_{\uparrow\downarrow}[\sig]^{}_{\uparrow\downarrow}+R^{L}_{\downarrow\uparrow}[\sig]^{}_{\downarrow\uparrow}+R^{L}_{\downarrow\downarrow}[\sig]^{}_{\downarrow\downarrow}\ .
\label{MR}
\end{align}
 Below we  calculate $I^L_{}$ and $\dot{\bf M}^L_{}$  for different orientations of $\hat{\bf z}$ relative to the link directions $\hat{\bf k},~\hat{\bf n}$ and $\hat{\bf d}$, see Fig. \ref{f1}.

The  rate Eq. (\ref{RA}) is found to second-order in the tunneling. 
Appendix A  generalizes the algebra of the Supplementary Material in Ref. \onlinecite{Aharony_2018}, ending up with  Eq. (\ref{RR}).
 Without any bias voltage, $\mu^{}_{L\sigma}=\mu^{}_{R\sigma'}=\mu$,  the two Fermi functions are equal to $f(\epsilon)=1/[e^{\beta(\epsilon-\mu)}+1]$, and then Eq. (\ref{RR}) yields $R^{L}_{\sigma\sigma'}=0$, so that the steady state particle current and all three components of the magnetization rate vanish, even when the densities of states do depend on the spin index. This result seems general, and valid for {\bf any} spin-dependent tunneling amplitude  $V^{}_{\bf kp}$ and any bare spin polarization in the leads.

Rewriting Eq. (\ref{RR}) as
\begin{align}
R^{L}_{\sigma\sigma'}&=2\pi\sum_{\sigma^{}_{1}}[V]^{}_{\sigma'\sigma^{}_1}[V^\ast]^{}_{\sigma\sigma^{}_1}\big\{{\cal N}^{}_{R,\sigma^{}_1}\big[N^L_{0\sigma}+N^L_{0\sigma'}\big]\nonumber\\
&-\big[{\cal N}^{}_{L\sigma}+{\cal N}^{}_{L\sigma'}\big]N^R_{0\sigma^{}_1}\big\}\ ,
\label{RR1}
\end{align}
and introducing the matrices
\begin{align}
\hat{N}^{R(L)}_{0}&=\left [\begin{array}{cc}N^{R(L)}_{0\uparrow}&0\\
0&
N^{R(L)}_{0\downarrow}\end{array}\right]
=\frac{1}{2}\Big (N^{R(L)}_{0}I+M^{R(L)}_{0z}\sigma^{}_{z}\Big )\ ,
\end{align}
where $I$ is the $2\times 2$ unit matrix and
\begin{align}
\hat{\cal N}^{}_{R(L)}=\left [\begin{array}{cc}{\cal N}^{}_{R(L)\uparrow}&0\\
0&
{\cal N}^{}_{R(L)\downarrow}\end{array}\right]\ ,
\end{align}
one finds
\begin{align}
I^{L}_{}&=4\pi{\rm Tr}\Big\{\hat{\cal N}^{}_{R}V^{\dagger}_{}\hat{N}^{L}_{0}V-\hat{N}^{R}_{0}V_{}^{\dagger}\hat{\cal N}^{}_{L}V\Big\}\ ,
\label{ILS}
\end{align}
and
\begin{align}
\dot{\bf M}^{L}_{}=&2\pi{\rm Tr}\Big \{\hat{\cal N}^{}_{R}V^{\dagger}_{}\Big (\hat{N}^{L}_{0}\sig +\sig \hat{N}^{L}_{0}\Big )V\nonumber\\
&-\hat{N}^{R}_{0}V_{}^{\dagger}\Big (\hat{\cal N}^{}_{L}\sig +\sig\hat{\cal N}^{}_{L}\Big )V\Big\}\ .
\label{MLS}
\end{align}
The last two equations  are valid  for any general tunneling matrix $V$.

\vspace{.5cm}

\section{Results for an SOI-active wire}\label{res}

We now  specialize to the unitary case, Eq. (\ref{VV}).
For this explicit form of the tunneling amplitude $V$,  one finds
\begin{widetext}
\begin{align}
&V^\dagger_{}\sig V=J^2 e^{i \alpha (\hat{\bf d}\cdot \sig)}\sig e^{-i\alpha (\hat{\bf d}\cdot\sig)}=J^2\big(\sig\cos(2\alpha)+\hat{\bf d}\times\sig \sin(2\alpha)+\hat{\bf d}(\hat{\bf d}\cdot\sig)[1-\cos(2\alpha)]\big)\ .
\label{rot1}
\end{align}
In this expression, the component of $\sig$ along $\hat{\bf d}$ remains unchanged, while the perpendicular component (which is in the $\hat{\bf n}-\hat{\bf k}$ plane) rotates  by an angle $2\alpha$ around the spin-orbit vector $\hat{\bf d}$, Eq. (\ref{ddd}).
Using this expression for $\sigma^{}_z$ also yields
\begin{align}
V^{\dagger}_{}\hat{N}^{L}_{0}V=\frac{J^{2}}{2}\Big [N^{L}_{0}+M^{L}_{0z}\Big (\sigma^{}_z\cos(2\alpha)
+[\hat{\bf z}\times\hat{\bf d}]\cdot\sig\sin(2\alpha)+2d^{}_{z}\sin^{2}\alpha(\hat{\bf d}\cdot\sig)\Big )\Big ]\ ,
\end{align}
where $d^{}_z=(\hat{\bf d}\cdot\hat{\bf z})$. It follows that
\begin{align}
&I^{L}_{}=2\pi J^{2}\Big \{N^{L}_{0}({\cal N}^{}_{R\up}+{\cal N}^{}_{R\down})-N^{R}_{0}({\cal N}^{}_{L\up}+{\cal N}^{}_{L\down})\nonumber\\
&+\Big [M^{L}_{0z}({\cal N}^{}_{R\up}-{\cal N}^{}_{R\down})-M^{R}_{0z}({\cal N}^{}_{L\up}-{\cal N}^{}_{L\down})\Big ]
 \Big [\cos(2\alpha)+2d^{2}_{z}\sin^{2}\alpha\Big ]\Big\}\nonumber\\
&=I^{L0}_{}+I^{L1}_{}(d^2_z-1) \sin^2\alpha
\ ,
\label{IILL}
\end{align}
\end{widetext}
where
\begin{align}
I^{L0}_{}&=4\pi J^2\big[{\cal N}^{}_{R\uparrow}N^L_{0\uparrow}+{\cal N}^{}_{R\downarrow}N^L_{0\downarrow}
-{\cal N}^{}_{L,\uparrow}N^R_{0\uparrow}-{\cal N}^{}_{L\downarrow}N^R_{0\downarrow}\big]
\end{align}
is the current into the left reservoir in the absence of the SOI, $\alpha=0$, and
\begin{align}
I^{L1}_{}= 4\pi J^2\Big [M^{L}_{0z}({\cal N}^{}_{R\up}-{\cal N}^{}_{R\down})
-M^{R}_{0z}({\cal N}^{}_{L\up}-{\cal N}^{}_{L\down})\Big ]\ .
\end{align}

To find the magnetization rate, we note that
\begin{align}
\hat{N}^L_0\sig+\sig\hat{N}^L_0=N^L_0\sig+M^L_{oz}\hat{\bf z}\ .
\end{align}
Similar algebra then yields
\begin{align}
\dot{\bf M}^L_{}=\dot{M}^L_z\hat{\bf z}+\dot{M}^L_\perp[\hat{\bf d}\times\hat{\bf z}]\ ,
\end{align}
with
\begin{align}
\dot{M}^L_z=\dot{M}^{L0}_{z}+2\dot{M}^{L1}_{\perp}(d^2_z-1) \sin^2\alpha
\ ,
\label{dMLz}
\end{align}
where
\begin{align}
\dot{M}^{L0}_{z}&=4\pi J^2\big[{\cal N}^{}_{R\uparrow}N^L_{0\uparrow}-{\cal N}^{}_{R\downarrow}N^L_{0\downarrow}
-{\cal N}^{}_{L\uparrow}N^R_{0\uparrow}+{\cal N}^{}_{L\downarrow}N^R_{0\downarrow}\big]
\end{align}
is the magnetization rate in the left reservoir in the absence of the SOI, and
\begin{align}
\dot{M}^{L1}_\perp=2\pi J^2\big[({\cal N}^{}_{R\uparrow}-{\cal N}^{}_{R\downarrow})N^L_{0}-({\cal N}^{}_{L\uparrow}+{\cal N}^{}_{L\downarrow})M^R_{0,z}\big]\ .
\end{align}
Also,
\begin{align}
\dot{M}^L_\perp=\dot{M}^{L1}_\perp\sin(2\alpha)\ .
\label{dMperp}
\end{align}

Consider first the case without the SOI, $\alpha=0$ [Eq. (\ref{Alp})]. Using the wide band approximation,  Eq. (\ref{NLRs}) becomes
\begin{align}
N^{L(R)}_{0\sigma}={\cal N}^{}_{L(R)\sigma}\int d\epsilon f^{}_{L(R)\sigma}(\epsilon)\ .
\label{NLRs1}
\end{align}
In the first scenario mentioned in Secs. \ref{INTR} and \ref{model}, the densities of states differ for the two spin components while  the chemical potentials (and therefore the Fermi functions) are spin-independent. In this case,
\begin{align}
I^{L0}_{}&=4\pi J^2\big[ {\cal N}^{}_{R\uparrow}{\cal N}^{}_{L\uparrow}+{\cal N}^{}_{R\downarrow}{\cal N}^{}_{L\downarrow}\big]\int d\epsilon[f^{}_L(\epsilon)-f^{}_R(\epsilon)]\ ,\nonumber\\
\dot{M}^{L0}_z&=4\pi J^2\big[ {\cal N}^{}_{R\uparrow}{\cal N}^{}_{L\uparrow}-{\cal N}^{}_{R\downarrow}{\cal N}^{}_{L\downarrow}\big]\int d\epsilon[f^{}_L(\epsilon)-f^{}_R(\epsilon)]\ .
\end{align}
Therefore,  currents will flow through the weak link only if  the density of states corresponding to at least one of the spin components does not vanish in {\it both} reservoirs. 
If the reservoirs are ferromagnetic half metals with opposite magnetizations then  e.g. ${\cal N}^{}_{L\downarrow}={\cal N}^{}_{R\uparrow}=0$, and then both charge and spin currents are fully blocked. As we show below, the SOI can open this blocking by flipping the spins in the weak link, as anticipated by Datta and Das.

In the second (non-equilibrium) scenario, the densities of states are spin-independent, and the Fermi functions depend on the spin components. In this case,
\begin{align}
I^{L0}_{}&=4\pi J^2{\cal N}^{}_L{\cal N}^{}_R\nonumber\\
&\times\int d\epsilon[f^{}_{L\uparrow}(\epsilon)+f^{}_{L\downarrow}(\epsilon)-f^{}_{R\uparrow}(\epsilon)-f^{}_{R\downarrow}(\epsilon)]\ ,\nonumber\\
\dot{M}^{L0}_z&=4\pi J^2{\cal N}^{}_L{\cal N}^{}_R\nonumber\\
&\times\int d\epsilon[f^{}_{L\uparrow}(\epsilon)-f^{}_{L\downarrow}(\epsilon)-f^{}_{R\uparrow}(\epsilon)+f^{}_{R\downarrow}(\epsilon)]\ .
\end{align}
These are the `usual' Landauer relations, in which electrons tunnel between the reservoirs without changing their polarizations.

When the reservoir polarizations are along the SOI vector, $\hat{\bf z}=\hat{\bf d}$ (i.e. $d^2_z=1$), then the SOI has no effect on the currents: $I^L_{}=I^{L0}_{}$ and $\dot{\bf M}^L_{z}=\dot{M}^{L0}_{z}\hat{\bf z}$.
 Since the spins of the electrons which move in the link rotate around the $\hat{\bf d}$ axis, it is not surprising that no magnetization is generated by a reservoir  magnetization directed along this rotation axis.
 For any other direction of the reservoir polarization, $d^2_z<1$, there may appear contributions to the currents which oscillate with $\alpha$, i.e. with the strength of the SOI and with the wire length [see Eq. (\ref{Alp})].\cite{flensberg} However, the amplitudes of these oscillating terms depend on the details of the reservoir polarizations, as reflected by the coefficients $I^{L1}_{}$ and $\dot{M}^{L1}_\perp$. As far as we know, this dependence has not yet been investigated in the literature.

In the first scenario we find 
\begin{align}
I^{L1}_{}=&4\pi J^2({\cal N}^{}_{R\uparrow}-{\cal N}^{}_{R\downarrow})({\cal N}^{}_{L\uparrow}-{\cal N}^{}_{L\downarrow})\nonumber\\
&\times\int d\epsilon[f^{}_L(\epsilon)-f^{}_R(\epsilon)]\ ,
\end{align}
and therefore the SOI-dependent oscillating term in the particle current vanished unless {\it both} reservoirs have spin-dependent densities of states, which means that the are {\it both} polarized.
If one of the reservoirs is not polarized then the particle current is not affected by the SOI. In the same case,
\begin{align}
\dot{M}^{L1}_\perp&=2\pi J^2({\cal N}^{}_{R\uparrow}-{\cal N}^{}_{R\downarrow})({\cal N}^{}_{L\uparrow}+{\cal N}^{}_{L\downarrow})\nonumber\\
&\times\int d\epsilon[f^{}_L(\epsilon)-f^{}_R(\epsilon)]\ .
\end{align}
A nonzero `bare' magnetization in the right reservoir generates a rotated polarization in the left reservoir, even when the `bare' magnetization on the left reservoir is equal to zero.  In particular, in the extreme Datta-Das case of oppositely polarized reservoirs, the `blocking' mentioned above opens, and both reservoirs will develop transverse oscillating magnetizations.
In the opposite case, when both reservoirs are fully polarized in the same direction, e.g.,  if ${\cal N}^{}_{L\downarrow}={\cal N}^{}_{R\downarrow}=0$, then \begin{align}
I^{L0}_{}&=I^{L1}_{}=\dot{M}^{L0}_z=2\dot{M}^{L1}_\perp\nonumber\\
&=4\pi J^2{\cal N}^{}_{L\up}{\cal N}^{}_{R\up}\int d\epsilon[f_L(\epsilon)-f_R(\epsilon)]\ .
\end{align}
and then both $I^L$ and $\dot{\bf M}^L$ oscillate with $2\alpha$, and vanish (i.e.,  both currents are blocked) when $d^2_z=0$ and $\sin^2\alpha=1$, again in agreement with Datta and Das.

In the second scenario the oscillating term in $I^L_{}$ vanishes, but
\begin{align}
\dot{M}^{L1}_\perp=-4\pi J^2{\cal N}^{}_L{\cal N}^{}_R\int d\epsilon[f^{}_{R\uparrow}(\epsilon)-f^{}_{R\downarrow}(\epsilon)]\ .
\end{align}
Again, the `bare' magnetization in the right reservoir generates a transverse rotating magnetization in the left lead (and vice versa).

 The currents into the right lead are found from the above equations by interchanging $L\Leftrightarrow R$. It is easy to check that the particle current is conserved,  $I^{L}_{}+I^{R}_{}=0$, as expected. The zero-SOI magnetization rates are also conserved, $\dot{M}^{L0}_{z}+\dot{M}^{R0}_{z}=0$: each electron tunnels keeping its spin unchanged. However, the oscillating part in the magnetization rates is generally not conserved,
 \begin{align}
 \dot{M}^{L1}_\perp+ \dot{M}^{R1}_\perp&=4\pi J^2\big[{\cal N}^{}_{R\up}N^L_{0\down}-{\cal N}^{}_{R\down}N^L_{0\up}\nonumber\\
 &+{\cal N}^{}_{L\up}N^R_{0\down}-{\cal N}^{}_{L\down}N^R_{0\up}\big]\ .
 \end{align}
Except for special cases, this remains nonzero. Therefore,
 the SOI induces a rotating spin current, which flows into the reservoirs.

\section{Summary and Conclusions}\label{sum}

In this paper we have investigated the effects of reservoir polarization on the currents through a spin-orbit active weak link.  Since our results depend on the relative directions of this polarization, which is along $\hat{\bf z}$, and of the SOI vector $\hat{\bf d}=\hat{\bf n}\times \hat{\bf k}$ (determined by the direction of the external electric field that tunes the Rashba SOI $\hat{\bf n}$ and by the direction of the weak link, $\hat{\bf k}$), rotating the electric field at a given polarization and varying its magnitude allows the measurement of the SOI strength in the wire: when $\hat{\bf z}=\hat{\bf d}$  there is no effect of the SOI. This effect grows gradually as $\hat{\bf d}$ is rotated away from $\hat{\bf z}$. The resulting particle current and magnetization rates oscillate as function of the SOI strength $\alpha$.

We considered two scenarios: in one, all the electrons have spin-independent chemical potentials, and the `bare' reservoir magnetizations result only from the spin-dependence of the reservoir densities of states.  In the other, the reservoir densities of state are spin-independent, but a microwave radiation can generate  non-equilibrium reservoir states which can be described by spin-dependent chemical potentials. In both scenarios,
The SOI is found to generate magnetization components in each reservoir, which rotate in the plane of the electric field (generating the SOI) and the weak link, only if the `bare' magnetization of the other reservoir has a non-zero component in that plane.
The SOI affects the charge current only in the first scenario, and only if both reservoirs are polarized.
 In both cases, the net charge current is conserved, but the transverse rotating magnetization current is not conserved; the SOI in the weak link generates extra spin polarizations which are injected into the reservoirs.

The SOI active weak link, schematically shown to connect two bulk electrodes in Fig.~\ref{f1}, could, e.g., be a nanowire comprising a semiconductor heterostructure or a quantum well. If the growth direction of a heterostructure is $\hat{\bf n}$, then the internal electric field across the heterostructure would produce an effective spin-orbit magnetic field along $\hat{\bf d}$ as assumed in our analysis. The strength of this effective field could be tuned by using electrostatic gates to apply in addition an external electric field along $\hat{\bf n}$. Such a tunability was demonstrated for an In$_{0.75}$Ga$_{0.25}$As/In$_{0.75}$Al$_{0.25}$As  heterostructure in Ref.~\onlinecite{Sato}. There, the spin-orbit coupling constant, mainly attributed to the Rashba SOI parameter $\alpha_R$, varied with gate voltage between roughly 150 and 300 meV \AA . Using $k_{\rm so} = m^\ast \alpha_R/\hbar^2$ and the measured $m^\ast = 0.041m$ ($m$ is the mass of a free electron), one concludes that with a 1$\mu$m-long weak link made of this material the parameter $\alpha=k_{\rm so}\ell$, which appears as an argument in the term $\sin^2\alpha$ in Eq.~(\ref{IILL}) for the current $I^L$ and  Eq.~(\ref{dMLz}) for the magnetization rate $\dot{M}^L_z$,
could be varied from $\sim8$ to $\sim 16$. This amounts to a tuning over a range that is more than twice the period $\pi$ of $\sin^2\alpha$. More recent experimental evidence for the SOI tunability is found in a dual gated InAs/GaSb quantum well where the Rashba SOI parameter $\alpha_R$ could be varied between 53 and 75 meV \AA , while the Dresselhaus SOI was kept constant. \cite{Beukman} The stated value of $m^\ast = 0.04m$ implies that $\alpha=k_{\rm so}\ell$ could be varied between $\sim$3 and $\sim$4, which is about a third of the period $\pi$ if $\ell = 1 \mu$m.

\vspace{.5cm}

\acknowledgments
This research was  partially supported by the Israel Science Foundation (ISF), by the infrastructure program of Israel Ministry of Science and Technology under contract 3-11173, and by the Pazy Foundation.

\appendix
\section{Calculating the rates}

To linear order in ${\cal H}^{}_{\rm tun}$ one finds [see
 the Supplementary Material in Ref. \onlinecite{Aharony_2018}]
\begin{widetext}
\begin{align}
&\langle c^{\dagger}_{{\bf k}\sigma}c^{}_{{\bf p}\sigma'}\rangle =i\int^{t}dt^{}_{1}\langle {\cal H}^{}_{\rm tun}(t^{}_{1})c^{\dagger}_{{\bf k}\sigma}(t)c^{}_{{\bf p}\sigma'}(t)-c^{\dagger}_{{\bf k}\sigma}(t)c^{}_{{\bf p}\sigma'}(t) {\cal H}^{}_{\rm tun}(t^{}_{1})\rangle
\nonumber\\
&=-i[V^{\ast}_{\bf kp}]^{}_{\sigma\sigma'}\Big (f^{}_{L,\sigma}(\epsilon^{}_{k})[1-f^{}_{R,\sigma'}(\epsilon^{}_{p})]-
f^{}_{R,\sigma'}(\epsilon^{}_{p})[1-f^{}_{L,\sigma}(\epsilon^{}_{k})]
\Big )
\int ^{t}dt^{}_{1}e^{i(\epsilon^{}_{k}-\epsilon^{}_{p}+i0^{+}_{})(t-t^{}_{1})}\nonumber\\
&=-[V^{\ast}_{\bf kp}]^{}_{\sigma\sigma'}\frac{
f^{}_{L,\sigma}(\epsilon^{}_{k})-
f^{}_{R,\sigma'}(\epsilon^{}_{p})}{
\epsilon^{}_{k}-\epsilon^{}_{p}+i0^{+}_{}}\ .
\end{align}

Substituting this expression in Eq. (\ref{RA}) yields
\begin{align}
R^{L}_{\sigma\sigma'}&=i\sum_{{\bf k},{\bf p},\sigma^{}_{1}}[V^{\ast}_{{\bf k}{\bf p}}]^{}_{\sigma\sigma^{}_{1}}
[V^{}_{\bf kp}]^{}_{\sigma'\sigma^{}_{1}}\Big \lbrace\frac{
f^{}_{L,\sigma'}(\epsilon^{}_{k})-
f^{}_{R,\sigma^{}_1}(\epsilon^{}_{p})}{
\epsilon^{}_{p}-\epsilon^{}_{k}+i0^{+}_{}}+\frac{
f^{}_{L,\sigma}(\epsilon^{}_{k})-
f^{}_{R,\sigma_1}(\epsilon^{}_{p})}{
\epsilon^{}_{k}-\epsilon^{}_{p}+i0^{+}_{}}\Big \rbrace\ .
\label{Rate11}
\end{align}
Assuming that the momenta dependence of $V_{\bf kp}$ and of the densities of states ${\cal N}^{}_{L,\sigma}$ and ${\cal N}^{}_{R,\sigma}$ is negligible (i.e., they are evaluated at the common average Fermi energy of the leads, in the so-called ``wide band approximation")
we may approximate
\begin{align}
R^{L}_{\sigma\sigma'}=\sum_{\sigma^{}_{1}}[V]^{}_{\sigma'\sigma^{}_{1}}[V^{\ast}_{}]^{}_{\sigma\sigma^{}_{1}}{\cal N}^{}_{R,\sigma^{}_{1}}
&\Big \{
{\cal N}^{}_{L,\sigma}\int dx\int dy[f^{}_{L,\sigma}(x)-f^{}_{R,\sigma^{}_{1}}(y)]\Big (i\frac{\cal P}{x-y}+\pi\delta(x-y)\Big )\nonumber\\
&+{\cal N}^{}_{L,\sigma'}\int dx\int dy[f^{}_{L,\sigma'}(x)-f^{}_{R,\sigma^{}_{1}}(y)]\Big (-i\frac{\cal P}{x-y}+\pi\delta(x-y)\Big )\Big \}\ .
\label{RLss}
\end{align}
Since $\int dx {\cal P}[1/(x-y)]=\int dy {\cal P}[1/(x-y)]=0$, the principal parts do not contribute, and therefore
\begin{align}
R^{L}_{\sigma\sigma'}&=\pi \sum_{\sigma^{}_{1}}[V]^{}_{\sigma'\sigma^{}_{1}}[V^{\ast}_{}]^{}_{\sigma\sigma^{}_{1}}{\cal N}^{}_{R,\sigma^{}_{1}}
\int dy \Big \{
{\cal N}^{}_{L,\sigma}[f^{}_{L,\sigma}(y)-f^{}_{R,\sigma^{}_{1}}(y)]
+{\cal N}^{}_{L,\sigma'}[f^{}_{L,\sigma'}(y)-f^{}_{R,\sigma^{}_{1}}(y)]\Big \}\nonumber\\
&=2\pi\sum_{\sigma^{}_{1}}[V]^{}_{\sigma'\sigma^{}_1}[V^\ast]^{}_{\sigma\sigma^{}_1}\big\{{\cal N}^{}_{R,\sigma^{}_1}\big[N^L_{0,\sigma}+N^L_{0,\sigma'}\big]-\big[{\cal N}^{}_{L,\sigma}+{\cal N}^{}_{L,\sigma'}\big]N^R_{0,\sigma^{}_1}\big\}\ .
\label{RR}
\end{align}
\end{widetext}



\end{document}